# Social Recommendations within the Multimedia Sharing Systems[1]

Katarzyna Musiał[1], Przemysław Kazienko[1] and Tomasz Kajdanowicz[2]

[1] Wrocław University of Technology, Wyb.Wyspiańskiego 27, 50-370 Wrocław, Poland
{katarzyna.musial, kazienko}@pwr.wroc.pl
[2] Hewlett Packard Polska Sp. z o.o., Warszawa, Poland
tomasz.kajdanowicz@hp.com

**Abstract.** The social recommender system that supports the creation of new relations between users in the multimedia sharing system is presented in the paper. To generate suggestions the new concept of the multirelational social network was introduced. It covers both direct as well as object-based relationships that reflect social and semantic links between users. The main goal of the new method is to create the personalized suggestions that are continuously adapted to users' needs depending on the personal weights assigned to each layer from the social network. The conducted experiments confirmed the usefulness of the proposed model.

**Keywords:** recommender system, multirelational social network, multimedia sharing system, social network analysis

## 1   Introduction

Nowadays, the multimedia sharing systems (MSS) like *Flickr* or *YouTube* successfully attract more and more users who share their multimedia content such as photos, videos, animations, etc. These systems that facilitate users to upload, download, manage and browse multimedia objects (MOs) are typical examples of Web 2.0 applications. Each of the multimedia object can be tagged by its author. In other words, users can describe their MO with one or more short phrases that are most meaningful for the authors and usually describe the content of this element. In the multimedia sharing system, users simultaneously interact, collaborate and influence one another and in this way form a kind of social community. Hence, users can not only tag multimedia objects they have published but also comment the items added by others, include them to their favorites, etc. Additionally, users have the opportunity to set up new, direct relationships with other system users as well as establish groups of collective interests and directly enumerate their friends or acquaintances.

---

[1] This is not the final version of this paper. You can find the final version on the publisher web page.

The main goal of every MSS is to enable people to share their MOs, i.e. that users usually contact with one another via MO and this is the basis for object-based relationship creation. Moreover, they very rarely maintain the direct personal relationships with other users of the system. Nevertheless, the users' activity within the MSS enables to discover people who act in a similar way. Based on that knowledge the direct social relationships can be set up what in consequence can lead to the development of the new dimension within the MSS, which enables people to contact with one another.

## 2 Related Work

Recommender systems became an important part of the web sites; the vast numbers of them are applied to e-commerce. They help people to make decision, what items to buy, which news to read [16] or which movie to watch. Recommender systems are especially useful in environments with information overload since they cope with selection of a small subset of items that appears to fit to the users' preferences [2, 11, 15, 18]. Furthermore, these systems enable to maintain the loyalty of the customers and increase the sales [10].

In general, three categories of recommender systems can be enumerated: collaborative filtering, content-based filtering, and hybrid recommendation [2]. The collaborative filtering technique relies on opinions about items delivered by users. The system recommends products or people that have been positively evaluated by other people, whose ratings and tastes are similar to the preferences of the user who will receive recommendation [2, 5, 16]. There are two main variants of collaborative filtering. The first one is the k-nearest neighbour and the second one is the nearest neighbourhood. In the content-based filtering the items that are recommended to the user are similar to the items that user had liked previously [12]. The hybrid method combines two previously enumerated approaches [7, 9, 10].

Nowadays, these three approaches are usually utilized in order to suggest different products or services to users. However, not only products or multimedia content can be proposed. The new area, where recommender systems can be applied are multimedia sharing systems that rapidly develop in the web and usually have thousands or even millions of members like *Flickr* or *YouTube*. The main goal of a recommender system in this case is to find the most interesting users for the given member and to help the user to establish new relationships.

The focus of this paper is to suggest one human being to another and in consequence to expand the human community that is not explicitly visible for users, because they are rather connected via MOs than direct links. The proposed recommendation framework is supposed to be applied to this community also called social network. A social network is the set of the actors (a single person is the node of the network) and ties, called also relationships, that link the nodes [1, 17]. The evolution of the social network depends on the mutual experience, knowledge, relative interpersonal interests, and trust of human beings [3, 14, 19]. The measurements can be collected to investigate the number and the quality of the relationships within the net-work. The crucial techniques currently used to identify the structure of a social network are: full network method, snowball method, and ego centric methods [6].

## 3 Multirelational Social Network in MSS

Based on the information about the MSS users and their activities, the multirelational social network (MSN) can be created. The network nodes are MSS users whereas the relationships emerge from the common activities or interaction between users. Overall, three kinds of relations can be distinguished in MSS: direct intentional relations, object-based relations with equal roles, and object-based relations with different roles. The first one exists if one user directly points to another one, e.g. by adding the given person to the contact list (Fig. 1).

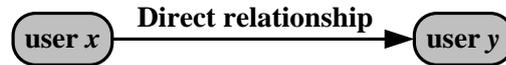

**Fig. 1** The direct tie in the social network in the Internet

The second type appears when two users participate in common activity related to the certain multimedia object (MO or tag) with the same role (Fig. 2a), e.g. users add to favourites the same MO or use the same tag. Finally, the object-based relation with different roles connects two users through the multimedia object but their roles are different (Fig. 2b), e.g. one user comments MO (*commentator*) that was published by another one (*author*).

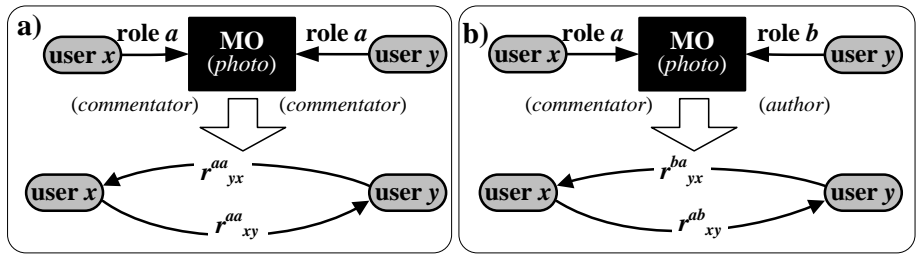

**Fig. 2** The object-based relation with equal roles: *commentator* (a), and different roles: *commentator* and *author* (b)

During the research, 11 types of relations were identified in the *Flickr* MSS, which can be classify in all the above kinds of relations (Fig. 3). They include: tags used by more than one user $R^t$, user groups $R^g$, MOs added by users to their favourites $R^{fa}$, $R^{af}$, $R^{ff}$, opinions about MOs created by users $R^{oa}$, $R^{ao}$, $R^{oo}$, and the relations derived from the contact lists $R^{cc}$, $R^{ac}$, $R^{cac}$. The contact-based direct relations can be split into three relations: user $x$ and $y$ are both in the contact list of another user $z$ ($R^{cc}$), $x$ is in the $y$'s contact list ($R^{ca}$), and $x$ is the $z$'s contact list but $z$ is also in the $y$'s contact list ($R^{cac}$). All the relations were used to create 11 layers in the multirelational social network. [13]

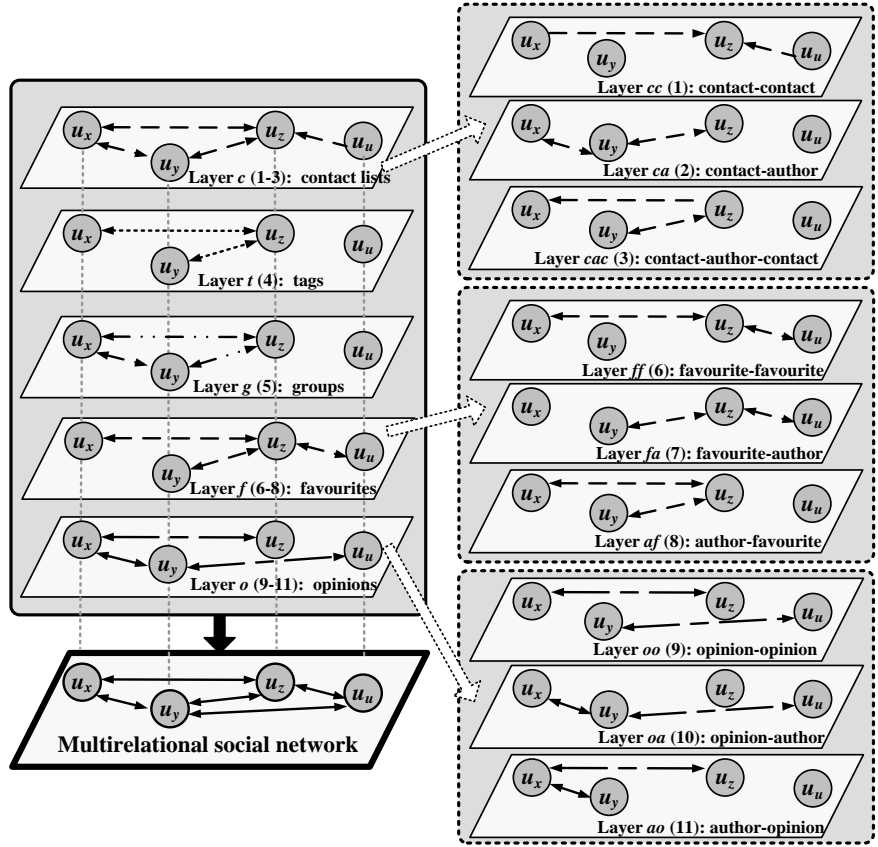

**Fig. 3.** The relation layers within the *Flickr* system

Tag-based $R^t$, group-based $R^g$, favourite-favourite $R^{ff}$, and opinion-opinion relations $R^{oo}$ are instances of object-based relations with equal roles, whereas favourite-author $R^{fa}$, author-favourite $R^{af}$, opinion-author $R^{oa}$, and author-opinion $R^{ao}$ are object-based relations with different roles. The appropriate strength value $s_{ikj}$ is calculated for each pair of users $k \rightarrow j$ and layer $i$.

## 4   Social Recommendations within MSS

The main idea of recommendations in the multimedia sharing system is to make use of relations between users that can be derived from the multirelational social network MSN existing in the MSS by recommendation to the active user some other users potentially interesting for the given one, Fig. 4.

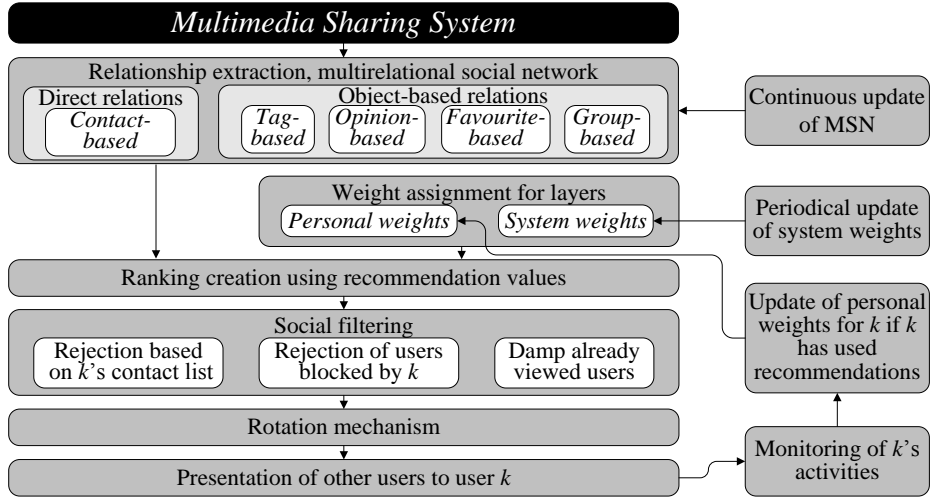

**Fig. 4** Recommendation of humans in MSS for user k using multirelational social network

In the first step, MSN is created and continuously updated based on the data available in the MSS, including new comments, MMOs, items in contact lists or groups, etc., see Sec. 2. MSN contains all 11 layers of relationships and provides similarities $s_{ikj}$ (strengths of relationships) from user $k$ to another user $j$, separately for each such pair $k \rightarrow j$ and each relationship layer $i$. The values of $s_{ikj}$ are calculated in the different way depending on the layer profile. The MSS system maintains two kinds of weights for each layer $i$ in MSN: system and personal. The system weight $w_i^{sys}$ for layer $i$ is the aggregation of all personal weights for layer $i$. It is updated periodically, e.g. ones a day. Personal weight of layers $w_{ik}^{usr}$ reflects the current usefulness of layer $i$ for the given user $k$. Both system and personal weights belong to the range [0;1] but the sum of $w_{ik}^{usr}$ for the given user $k$ equals 1. For the new user $k$, $w_{ik}^{usr}=w_i^{sys}$ is assigned. All personal weights for user $k$ are updated according to $k$'s activities that refer to the recommended persons $j$ like browsing $j$th profile, adding $j$ to the $k$th contact list, comments to $j$th MMOs, etc. In the experimental environment, users were requested to rate the presented recommendations and these rates were used as the feedback from user activities.

Based on the similarities derived from MSN as well as system and personal weights, the system calculates the recommendation values $v_{kj}$ for the current user $k$ related to the other users $j$, as the aggregations of similarities from all $l$ layers:

$$v_{kj} = \sum_{i=1}^{l} \frac{(w_i^{sys} + w_{ik}^{usr}) \cdot s_{ikj}}{\max_i (s_{ikj})}. \qquad (1)$$

The recommendation values $v_{kj}$ are used to create the ranking list for user $k$ that contains top users $j$ with the greatest value of $v_{kj}$. Next, some users $j$ are removed during the social filtering process. Its goal is to prevent from recommendation of users that already are in the $k$th contact list or are blocked by $k$. Besides, the recommendation values of users who have already been viewed by $k$ are damped. To the remaining list rotation mechanism is applied so that the recommendation list changes with every user request to the system, see [8] for details. After presentation, the sys-tem monitors activities of user *k* related to the recommended users *j*. It includes viewing the *j*'s profile and establishment of the new relation $k \to j$ in any layer. The level of interest of user *k* directed to *j* reflected by *k*'s activities can be lower (viewing the *j*th profile) or greater (adding to *k*'s contact list). Hence, each type of activity possesses its own importance $a_{kj} \in [0;1]$. Next, based on this feedback, *k*'s personal weights $w_{ik}^{usr}$ are adapted after each *k*'s relevant activity, as follows:

$$w_{ik}^{usr(new)} = \frac{w_{ik}^{usr(old)} \cdot (1+\varepsilon) + c_{ikj} \cdot \left(a_{kj} - w_{ik}^{usr(old)}\right)}{\sum_{m=1}^{l}\left(w_{mk}^{usr(old)} + c_{mkj} \cdot \left(a_{kj} - w_{mk}^{usr(old)}\right)\right)}, \quad (2)$$

where: $\varepsilon$ is a small constant; $c_{ikj} \in [0;1]$ is the normalized contribution of the *i*th layer (among all layers) within the recommendation of user *j* to user *k*; $c_{ikj} = s_{ikj}/\sum_{m=1}^{l} s_{mkj}$. Equation (2) preserves property of $w_{ik}^{usr(new)}$ auto-balancing in the range of [0;1] and takes into account the perspective on the global importance of particular kinds of relations in the entire MSN.

## 5   Experiments

The experiments have been carried out based on the online *FlickrFront* framework. During the experiments, 21,640 user profiles were downloaded from *Flickr* to prepare recommendations for 8 selected users who rated the provided suggestions. The rates replaced the monitored user activities $a_{kj}$ in Eq. (2). Upon the collected profiles and the concept from sec. 3, two separate recommendation lists, three suggested users each, were presented to 8 users who rated them. The adaptation derived from users' ratings ($a_{kj}$) and layer contributions was applied after the first list was rated.

Users have generally rated higher the recommendations provided in the second list (after adaptation), Fig. 5a. Besides, the social layer based on contact list $R^{cac}$ and MO author-opinion $R^{ao}$ gained much after adaptation, tag-based $R^t$ increased a little while the other lost in their importance, Fig. 5b. The least vital layers are $R^{oa}$, $R^{ff}$, $R^{af}$.

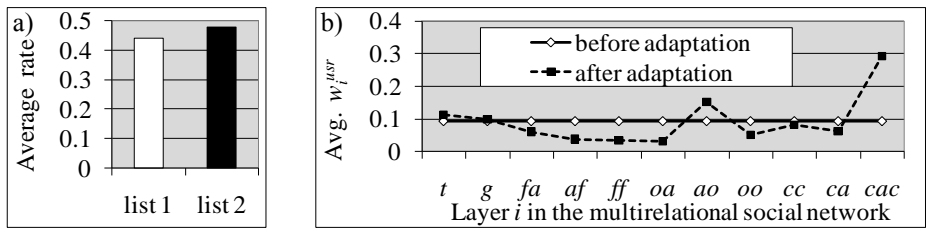

**Fig. 5** Average users' ratings (a) and average user weights for layers in MSN (b)

## 6   Conclusions and Future Work

The new proposed method supports the creation of recommendations of humans in the multimedia sharing system MSS based on the new social concept – the multire-lational social network MSN. The presented framework takes into account the activities of users in each of eleven MSN layers extracted from the MSS data. Both system and personal weights that are assigned separately to each layer make the process of recommendation personalized. Moreover, the system is adaptive due to weights that are adaptively recalculated when the user utilizes the recommendations. The vast amount of calculations results in problems with efficiency as the whole process is performed online. In order to address this issue, some tasks can be performed offline and periodically repeated, e.g. the creation of the lists and storing only $n$ most similar users to the given one. The future work will concentrate on improving the process effectiveness and conducting the large-scale experiments.